\documentclass[aps,twocolumn,pre,floatfix]{revtex4}
\usepackage{graphicx}

\newcommand{\br}{{\bf r}}

\begin{document}

\title{Density fluctuations of polymers in disordered media}

\author{J.M. Deutsch}
\affiliation{Department of Physics, University of California, Santa Cruz CA 95064}

\author{M. Olvera de la Cruz}

\affiliation{Department of Material Science and Engineering, Northwestern University, Evanston IL 60201}

\date{\today}

\begin{abstract}
We study self avoiding random walks in an environment where sites are excluded
randomly, in two and three dimensions. For a single polymer chain, we study the
statistics of the time averaged monomer density and show that these are well
described by multifractal statistics. This is true even far from the percolation
transition of the disordered medium. We investigate solutions of chains in a
disordered environment and show that the statistics cease to be multifractal
beyond the screening length of the solution.
\end{abstract}

\maketitle

\section{Introduction}

In many experimental situations, linear polymers are present in a disordered
medium. For example, a free polymer chain inside an elastomeric network, or
a DNA molecule in an agarose gel. The properties of single chains in such
situations have been the subject of numerous theoretical and numerical
work~\cite{CatesBall,MachtaGuyer,DoussalMachta,LeeNakanishi,Obukhov,GersappeDeutschOlvera,JanssenStenull,SungYethiraj,SungChangYethiraj}.
From a theoretical perspective, the problem is of great interest
in part because of its strong connection to the phenomenon of quantum
mechanical localization~\cite{Anderson}. The greens function for a 
single ideal polymer chain in a potential is mathematically identical
to that of an electron in the same potential, but in imaginary
time~\cite{DeGennesBook}. Therefore a polymer chain is expected to
collapse to a small region in space. However the presence of excluded
volume of the chain prevents this from occurring. The statistics of
a chain in this case was the subject much controversy. This was settled
by the work of Cates and Ball~\cite{CatesBall} who showed that the
statistics of a chain on an infinite lattice with frozen disorder,
that is the quenched case, was identical to that of a chain with
mobile defects, that is the annealed case. They showed this by dividing
up the infinite lattice into large finite regions with frozen disorder. The
average properties of a chain are obtained by averaging over all these large
finite regions. This is equivalent to performing an annealed average.
It is easy to perform the annealed average over uncorrelated random disorder and show
that this has no effect of chain statistics. Simulations~\cite{LeeNakanishi} away from
the percolation threshold bear out this prediction. Work on off-lattice
models of spherical obstacles show more complicated behavior, because
the correlations in the disorder are no longer point-like, so this
influences the conformation of a chain. For low obstacle volume fraction,
these correlations induce an effective attraction causing a decrease
in chain size. As the volume fraction gets close to the percolation
threshold~\cite{SungYethiraj,SungChangYethiraj}, the exclusion of phase
space for finite lattice sizes leads to an increase in the average chain size.

Although chain statistics in the above situation are not affected by
frozen disorder, there is another important physical quantity that
does change. Consider the density of monomers for a given background
of disorder. If this is averaged over time, one expects that the
density will fluctuate because of the fluctuations in the random
potential. This problem was studied numerically by Gersappe et al~\cite{GersappeDeutschOlvera} in
two dimensions~\cite{GersappeDeutschOlvera} where it was shown this time
averaged monomer density (TAMD) obeyed multifractal statistics~\cite{Mandelbrot,Benzi,Halsey} over the
parameter range studied. The spectrum of multifractal dimensions was
computed numerically and gave strong evidence for multifractal scaling.

More recently~\cite{BlavatskaJanke}, the properties of a polymer chain
on the backbone of a percolating cluster were examined numerically at
the percolation transition. They found that it exhibited multifractal
properties in agreement with the theoretical work of Janssen and
Stenull~\cite{JanssenStenull}. A measure of bond density for an ensemble
of chains of varying lengths that connect two points separated by a
distance $R$ was calculated numerically. That measure was analyzed and
shown to have a spectrum of multifractal exponents.

The earlier work of Gersappe et al~\cite{GersappeDeutschOlvera} in two dimensions claimed that multifractal statistics
were true even far from the percolation transition where the disorder is
dilute. In other words, the multifractal nature of polymers in disordered
systems is much more general and in fact, a simpler more experimentally
accessible quantity, namely the time averaged monomer density, can be used as
a measure.

Multifractal distributions are characterized by very large fluctuations,
so in this case, we expect that the TAMD $\rho(\br)$ will have a probability
distribution that becomes increasingly broad as the system size in increased.
We also expect multifractal statistics to be present for scales less than
the average radius of gyration, $R_g$, of the chain. On larger scales, the density fluctuations
should saturate so that the distribution $P$ of $\rho$ should obey~\cite{Benzi,Halsey}
\begin{equation}
\label{eq:falpha}
P(\log \rho) \propto \exp(\log(R_g) f(\log(\rho)/\log(R_g)) .
\end{equation}

In this paper, we study this problem in two dimensions for different amounts
of disorder and show that multifractality is seen in all cases. The original
work of Gersappe et al only presented results for one value of disorder. Next
we study this system in three dimensions and show that multifractal statistics
persist for this case as well. Next we consider many chains at finite concentrations
and demonstrate how this smooths out density fluctuations so that
the statistics are no longer multifractal.

In section \ref{sec:model} we describe the model that we will apply to study this
problem. In section \ref{sec:singlechains2d} we study the TAMD in two dimensions.
The results are similar to what was found earlier~\cite{GersappeDeutschOlvera} but
with much improved data providing more solid evidence for multifractality of a range
of obstacle densities. Section \ref{sec:singlechains3d} considers the same problem in
three dimensions. Although box sizes are smaller, this work provides solid support
for the case the TAMD is also a multifractal measure in this case as well. In
section \ref{sec:manychains3d} this problem is analyzed for a solution of chains
at finite concentration where it is clear that the system ceases to be multifractal
over the scale of the radius of gyration of a chain. An interesting bimodal distribution
for the distribution of the TAMD is found.

\section{The Model}
\label{sec:model}

We consider a cubic lattice in $d = 2$ and $3$ dimensions, with $m$ self
avoiding walks of unit step length of $N$ monomers. The monomers
move in a cubic box of width $L$. Skew boundary conditions are
employed~\cite{skew}. Obstacles are randomly placed on sites which are
excluded from occupation by polymer chains. We denote the average
fraction of excluded sites as $n_o$.

The chains are moved using ``reptation dynamics"~\cite{wallmandell}.
Briefly, the head or chain of a chain is picked at random and an attempt
is made to move it to the other end. The attempt fails if the site
is occupied by a chain or an obstacle, otherwise it succeeds. After
equilibrating, statistical properties of the chains are measured, such
as the radius of gyration and the time averaged monomer density.

To test this method, we checked that the internal chain statistics of
a single chain with random disorder were identical within statistical
error to that of a self avoiding walk with no disorder.

Next we considered density fluctuations.
Moments of the normalized time averaged monomer density $\rho_r$ can be calculated
by summing moments of the density over all lattice sites and inferring how
this depends on lattice size
\begin{equation}
\langle \sum_r^{L^d} \rho_r^q \rangle \sim L^{-\tau(q)}
\end{equation}
The average is over different realizations of the disorder.
However, it is more accurate to obtain the exponents $\tau(q)$ by coarse
graining, making use of the self-similar nature of multifractals.
If we coarse grain over boxes of width $l$, and define ${\tilde \rho}_r$ as
the density measure coarse grained over that length scale, then it is straightforward
to show that 
\begin{equation}
\label{eq:coarsetaudef}
T_q(l) \equiv \langle \sum_r^{(L/l)^d} {\tilde \rho}_r^q \rangle \sim l^{\tau(q) - d q} .
\end{equation}

By analyzing how this sum depends on $l$, $\tau(q)$ can then be determined.

\section{Single chains in two dimensions}
\label{sec:singlechains2d}

\begin{figure}[htp]
\begin{center}
(a)\includegraphics[width=\hsize]{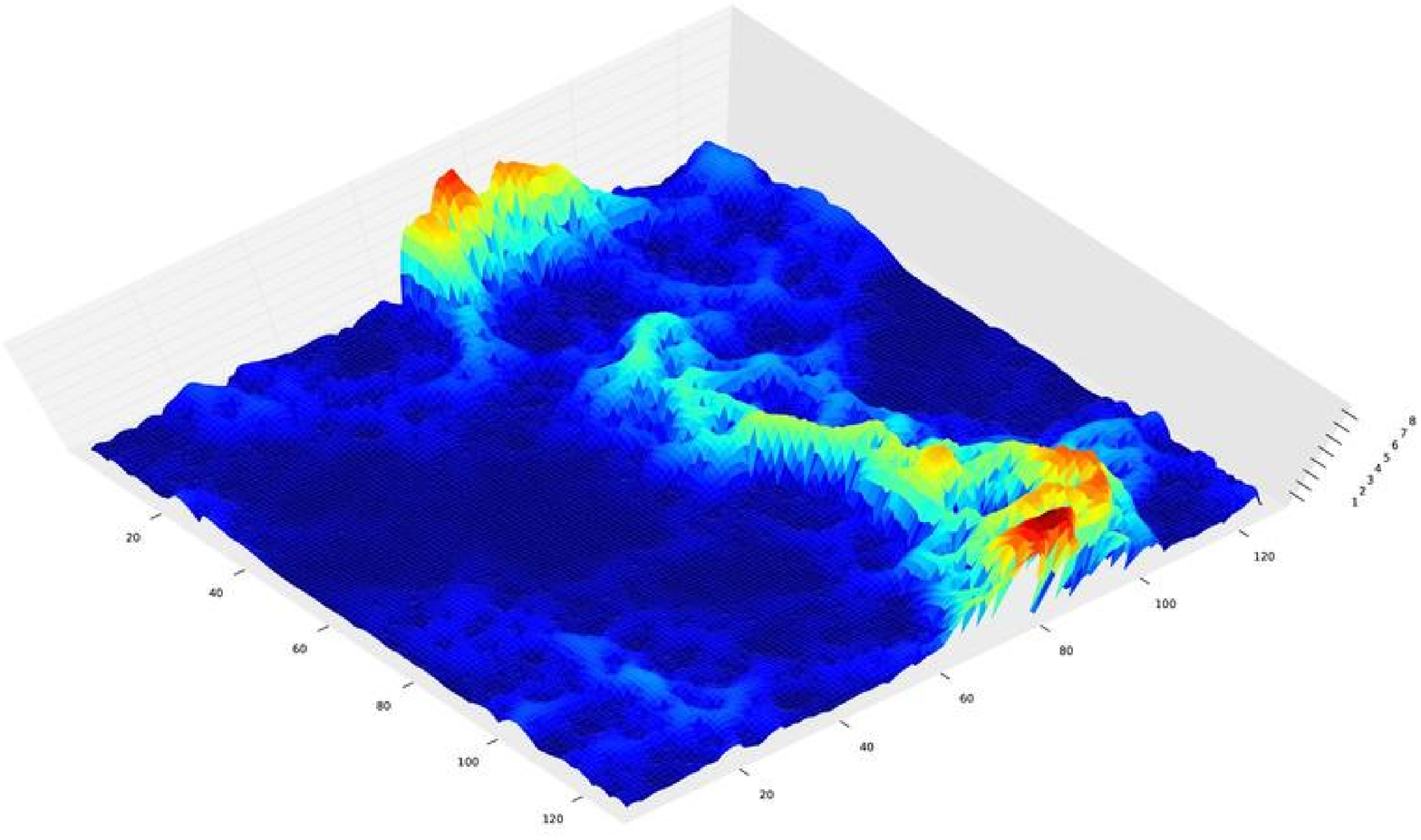} \\
(b)\includegraphics[width=\hsize]{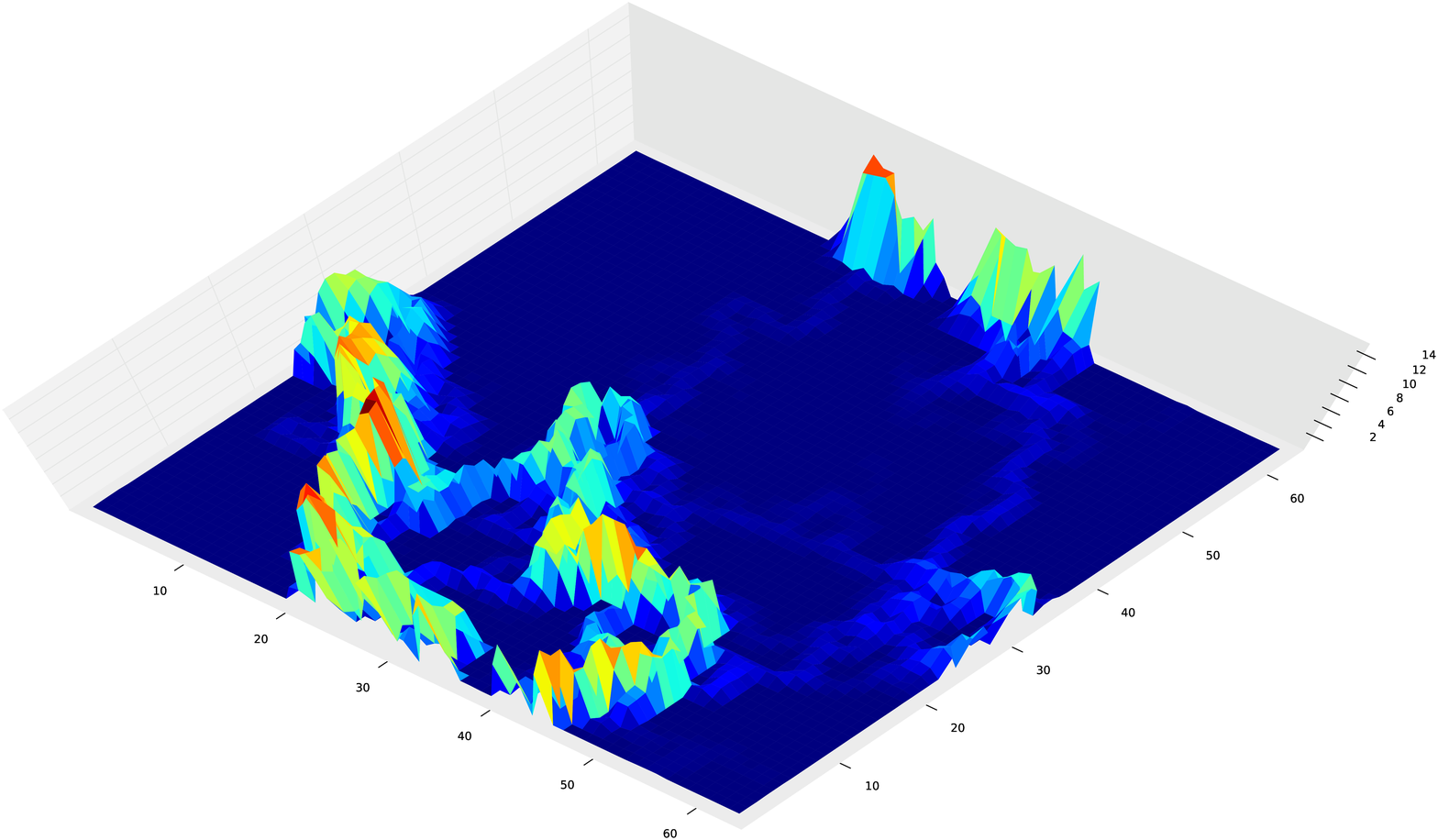}
\caption{ (Color Online)
The time averaged monomer density for a self avoiding random walk in two
dimensions, (a) $N=512$, $L=128$, and $n_o = 0.1$;
(b) $N=256$, $L=64$, and $n_o = 0.2$.
}
\label{fig:2dTAMD}
\end{center}
\end{figure}

The above model was studied for a single chain in two dimensions for
different values of the disorder and chain lengths. The TAMD $\rho_r$
is plotted in Figure \ref{fig:2dTAMD}.  Figure \ref{fig:2dTAMD}(a) shows
a chain of length $N=512$ in a box of length $L=128$ with an obstacle 
density of $n_o = .1$. The self similar nature
of the measure is apparent. 
Figure \ref{fig:2dTAMD}(b) shows the TAMD for $N=256$ and $L=64$, with $n_o = 0.2$.

The density shows a self similar structure with different regions appearing
statistically quite similar but differing by an overall multiplicative constant
as expected for a multifractal measure.  For distances larger than
the radius of gyration, one expects multifractal scaling to no longer hold. Similarly, for distances smaller
than some cut-off distance $l_0$, related to the density of obstacles, the density should become smooth
so that multifractal scaling no longer applies. But over intermediate scales, as can be seen,
the density appears self similar. This will be now quantified below.

\begin{figure}[htp]
\begin{center}
\includegraphics[width=\hsize]{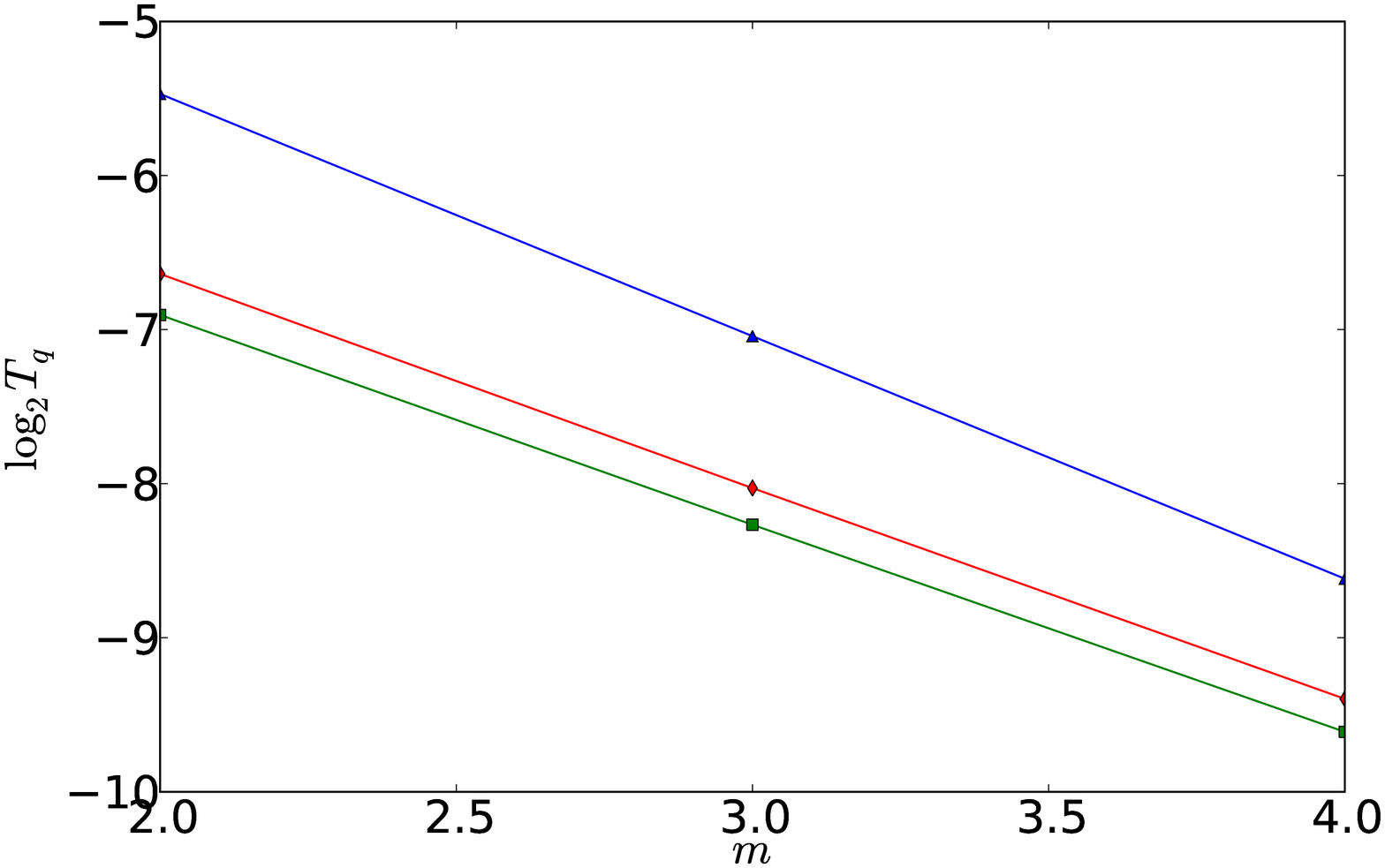} 
\caption{ (Color Online)
The scaling of moments of the coarse grained density $T_q$ as defined in Eq. \ref{eq:coarsetaudef}
as a function of the amount of coarse graining $m$ in two dimensions. The top line is data for $q=-1$,
the middle, $q=2$, and the bottom is for $q=1$.
}
\label{fig:2dscaling}
\end{center}
\end{figure}

Coarse graining of $\rho$  is done by recursive decimation of the
original lattice into $2\times 2$ blocks. This implies that the coarse graining length 
is a power of $2$, $l = 2^m$ where $m$ is the number of times the lattice is decimated. 
Good fits to power laws were obtained between the lower and upper cutoffs for Eq. \ref{eq:coarsetaudef} as
shown in Fig. \ref{fig:2dscaling}.
This plots the left hand side of Eq. \ref{eq:coarsetaudef} as a function of coarse graining
$m = \log_2 l$. Fig. \ref{fig:2dscaling} shows the case $N=512$, $L=128$, and $n_o = 0.2$  for $q = -1, 0, 1, 2$.
The results were averaged over $200$ different realizations
of the random obstacles, and each simulation was run for a total of $8\times 10^9$
steps. 
This is strong evidence of self similar multifractal behavior {\em far away} from the percolation
threshold in two dimensions.

\begin{figure}[htp]
\begin{center}
\includegraphics[width=\hsize]{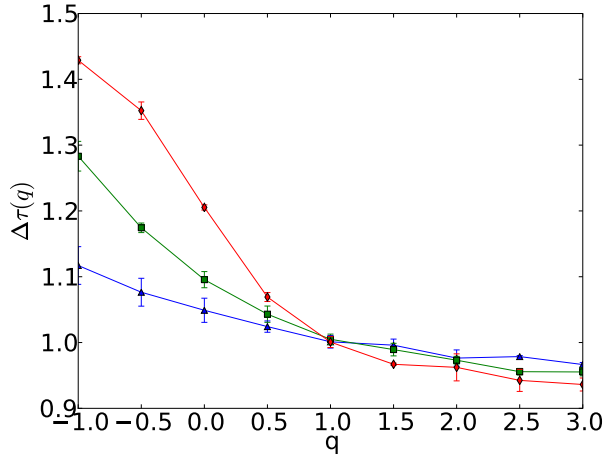} 
\caption{ (Color Online)
The results of simulations to determine the multifractal dimensions for
for different obstacle densities $n_0 = 0.1$ (triangles), $0.2$ (squares), and $0.3$ (diamonds) in two dimensions.
}
\label{fig:2dtauq}
\end{center}
\end{figure}

Using these fits, $\tau(q)$ defined in Eq. \ref{eq:coarsetaudef} was calculated.
The results are shown in Fig. \ref{fig:2dtauq}. For clarity we plot the difference
in $\tau$ defined as $\Delta\tau(q) \equiv \tau(q)-\tau(q-1/2)$. This is shown
for three different sets of parameters, $n_0 = 0.1$, $0.2$, and $0.3$.

The data give strong evidence for multifractal behavior for a range of different
obstacle densities.

\section{Single chains in three dimensions}
\label{sec:singlechains3d}

\begin{figure}[htp]
\begin{center}
\includegraphics[width=\hsize]{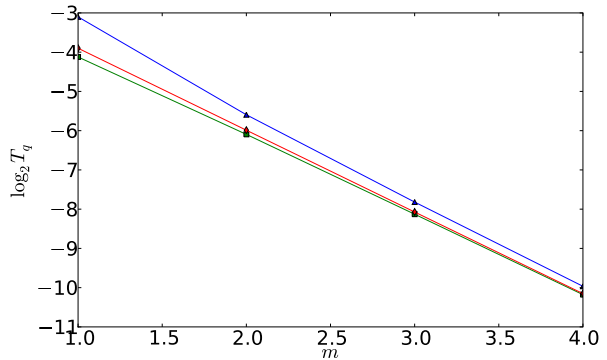} 
\caption{ (Color Online)
The scaling of moments of the coarse grained density $T_q$ as defined in Eq. \ref{eq:coarsetaudef}
as a function of the amount of coarse graining $m$ in three dimensions for $n_0 = 0.3$. 
The top line is data for $q=-1$, the middle, $q=2$, and the bottom is for $q=1$.
}
\label{fig:3dscaling}
\end{center}
\end{figure}

\begin{figure}[htp]
\begin{center}
\includegraphics[width=\hsize]{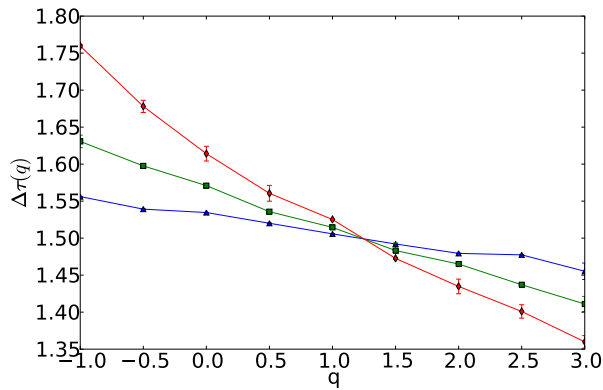} 
\caption{ (Color Online)
The results of simulations to determine the multifractal dimensions for
for different obstacle densities $n_0 = 0.1$ (triangles), $0.2$ (squares), and $0.3$ (diamonds) in three dimensions.
}
\label{fig:3dtauq}
\end{center}
\end{figure}

The simulation was run in three dimensions and analyzed for the possibility of multifractal scaling,
following the same procedure as in two dimensions. We first
plot $\langle {\tilde \rho}^q\rangle$
as a function of the coarse grained length scale $l$, as is defined in
relation to Eq. \ref{eq:coarsetaudef}. Fig. \ref{fig:3dscaling} shows the case $N=128$, $L=32$, and $n_o = 0.3$.
Because the box size and the chain length are smaller than in two dimensions, power law scaling can only
be seen over a more limited range, however, it does support the hypothesis that the TAMD is multifractal.
For $q=2$ a slight curvature can be seen in the log-log plot which is to be expected from finite size
effects. Overall the fits to power laws over this range is very good. Similarly good fits are also seen
for $n_0 = 0.2$. In Fig. \ref{fig:3dtauq}, we display fits for $\Delta \tau$ as a function of $q$ for $n_0 = 0.1$, $0.2$, and $0.3$.

Therefore is appears that in  three dimensions, our simulations  and
analysis also support the idea that the TAMD is multifractal for obstacle
densities away from the percolation.

\section{Many chains}
\label{sec:manychains3d}

The internal statistics of a self avoiding walk (SAW) in a much larger lattice of quenched random
obstacles was shown to be equivalent to that of an annealed average and hence that of a SAW without disorder~\cite{CatesBall}.
For the case of a many chain solution, extending this result is not completely straightforward.
The difficulty in this case is that the polymer solution extends throughout the entire lattice.
However, this problem is easily amenable to numerical simulation. Various equilibrium internal averages
for chains were calculated and found to be almost entirely unaffected by the disorder. As an example,
we considered $m = 40$ chains each of
length $N=128$ with an obstacle density of $n_0 = 0.2$ in a box of
size $L=32$. Excluding obstacles, this corresponds to a chain filling
fraction of $20\%$.  $R_g^2 = 49.190 \pm 0.013$
in this case. In comparison $R_g^2 = 48.785 \pm 0.006$ when run with no obstacles. The
difference between these two cases is less than $1\%$. As the obstacle density approaches
the percolation point, one would expect the difference between these two cases to increase 
due to the increasing likelihood of islands that are inaccessible, similar to the single
chain case~\cite{LeeNakanishi,SungYethiraj}.

In contrast to internal chain statistics, such as the average radius of gyration, we expect that 
the TAMD for the many chain case to be greatly affected by the presence of disorder,
as it was for the single chain case.

For a system of chains at finite density in three dimensions, the
fluctuations in chain density will be quite different than for a single
chain. Above the correlation length, correlations in density will decrease
rapidly. Therefore the distribution of the TAMD is no longer expected
to be multifractal over these scales. This problem was simulated using
the same approach as above.

\begin{figure}[t]
\begin{center}
\includegraphics[width=\hsize]{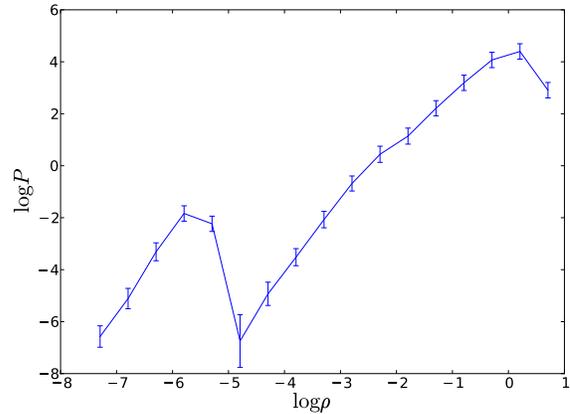} 
\caption{ (Color Online)
The distribution of the TAMD $\rho$ for $40$ chains,
an obstacle density of $n_0 = 0.2$ in three dimensions with $N=128$, $L=32$.
}
\label{fig:rhohist}
\end{center}
\end{figure}

The distribution of the TAMD was averaged over lattice sites and
different realizations of the obstacle density in three dimensions.
This is displayed in Fig. \ref{fig:rhohist} for the same system as discussed above:
$m = 40$ chains each of length $N=128$ with an obstacle density of $n_0 = 0.2$ in a box of
size $L=32$. The distribution is plotted with logarithmic axes
(base $e$). The results were averaged over $155$ different realizations
of the random obstacles, and each simulation was run for a total of $8\times 10^9$
steps. The TAMD $\rho$ was normalized so that $\langle \rho \rangle = 1$. 

The results show that the distribution $P(\log \rho)$ has two peaks.
One peak is close to $\rho =1$ and most values of the density are clustered
around that value. However, there is another peak for very small values of
the obstacle density $\rho = 0.003$. The second peak is unexpected because
it implies that there are intermediate values of the density that are much
less probable. 

The same feature persists when the obstacle density $n_0$ is reduced to $0.1$ and 
in both cases it is clear that this second peak is not due to random statistical
fluctuations. This is likely due to the presence of some particular local configurations
of obstacles, for example a cul-de-sac. In this case there will be entropic exclusion
of chain for such a region leading to a lower monomer density. 

The distribution in Fig. \ref{fig:rhohist} is is not compatible with
the prediction of multifractal statistics (Eq. \ref{eq:falpha}) as
the distribution must be convex. In addition, the coarse graining analysis of the last
two sections does not yield a spectrum of exponents. $\Delta \tau(q)$ over the same range
as Fig. \ref{fig:3dtauq} varies by less than $ 0.05$. The lack of large fluctuations is to be
expected due to the presence of a screening length. Although it is likely
that the TAMD is multifractal below this scale, it is not possible to
investigate this numerically because the accessible sizes only cover a
small range in scales.

\section{Acknowledgments}

We acknowledge was the support of the Non-equilibrium Energy Research Center (NERC) which is an Energy Frontier Research Center funded by the U.S.
Department of Energy, Office of Science, Office of Basic Energy Sciences under Award Number DE-SC0000989.


\begin{thebibliography}{}
\bibitem{CatesBall} M. E. Cates and C. Ball, J. Phys. (France) {\bf 49}, 2009 (1988).
\bibitem{MachtaGuyer} J. Machta and R.A. Guyer, J. Phys. A {\bf 22}, 2539 (1989)
\bibitem{DoussalMachta} P. Le Doussal and J. Machta, J. Stat. Phys. {\bf 64}, 541 (1991).
\bibitem{LeeNakanishi} S. B. Lee and H. Nakanishi, Phys. Rev. Lett. {\bf 61}, 2022 (1988) and reference therein.
\bibitem{Obukhov} S. P. Obukhov, Phys. Rev. A. {\bf 42B}, 2015  (1990).
\bibitem{JanssenStenull} H.K. Janssen and O. Stenull, Phys. Rev. E {\bf 75}, 020801(R) (2007).
\bibitem{GersappeDeutschOlvera} D. Gersappe, J.M. Deutsch and M. Olvera de la Cruz, Phys. Rev. Lett. {\bf 66} 731 (1991).
\bibitem{SungYethiraj} B. J. Sung and A. Yethiraj J. Chem. Phys. {\bf 123}, 074909 (2005).
\bibitem{SungChangYethiraj} B. J. Sung, R. Chang, and A. Yethiraj, J. Chem. Phys. {\bf 130}, 124908  (2009).
\bibitem{Anderson} P.W. Anderson Phys. Rev. {\bf 109}, 1492 (1958). 
\bibitem{DeGennesBook} P.G. de Gennes ``Scaling Concepts in Polymer Physics" Cornell University Press (1985).
\bibitem{Mandelbrot} B. Mandelbrot, ``Statistical Models and Turbulence", La Jolla, California, 1972. Edited by Murray Rosenblatt and
Charles Van Atta, New York, Springer. (Lecture Notes in Physics, 12), page 331 (1972).
\bibitem{Benzi} R. Benzi, G. Paladin, G. Parisi, and A. Vulpiani, J. Phys. A. {\bf 18}, 3521 (1984). 
\bibitem{Halsey} T.C. Halsey, M.H. Jensen, L.P. Kadanoff, I. Procaccia, and B.E. Shraiman, Phys. Rev. A {\bf 33} 1141 (1986).
\bibitem{BlavatskaJanke} V. Blavatska and W. Janke, Phys. Rev. Lett. {\bf 101} 125701 (2008).
\bibitem{skew} T.A. DeGrand and D. Toussaint, Phys Rev D {\bf 22} 2478 (1980).
\bibitem{wallmandell} F.T. Wall and F. Mandell, J. Chem. Phys. {\bf 63} 4592 (1975).
\end{thebibliography}
\end{document}